\def\nn#1 #2{#2. #1}				
\def\nnn#1 #2 #3{#2. #3. #1}			
\def\nnnn#1 #2 #3 #4{#2. #3. #4 #1}		
\def\nnnnn#1 #2 #3 #4 #5{#2. #3. #4 #5. #1}	

\def\pp{\noindent\parshape 2 0truecm 13.6truecm 1truecm 12.6truecm}
\def\rl#1;#2;#3;#4;#5 {\addtocounter{enumi}{1}\item[{[\arabic{enumi}]}]\par``#1", #2, {\it #3}, {\bf #4}, #5 \par}
\def\rlnonref#1;#2;#3;#4;#5 {\addtocounter{enumi}{1}\item[{ [\arabic{enumi}]}]\par``#1", #2, {\it #3}, {\bf #4}, #5 \par}
\def\pop#1;#2;#3; {\addtocounter{enumi}{1}\item[{[\roman{enumi}]}]\par``#1", {\it #2}, #3 \par}

\def\rf#1;#2;#3;#4 {\par\pp#1, {\it #2}, {\bf #3}, #4. \par}
\def\rk#1;#2;#3;#4;#5 {\par``#1", #2, {\it #3}, {\bf #4}, #5 \par}
\def\ro{}

\def\beq#1{\begin{equation}\label{#1}}
\def\eeq{\end{equation}}
\def\beqa#1{\begin{eqnarray}\label{#1}}
\def\eeqa{\end{eqnarray}}

\def\fig#1{Figure~\ref{#1}}

\def\tabl#1{Table~\ref{#1}}

\def\ie{{\frenchspacing\it i.e.}}
\def\eg{{\frenchspacing\it e.g.}}
\def\etc{{\frenchspacing\it etc.}}

\def\spose#1{\hbox to 0pt{#1\hss}}
\def\simlt{\mathrel{\spose{\lower 3pt\hbox{$\mathchar"218$}}
     \raise 2.0pt\hbox{$\mathchar"13C$}}}
\def\simgt{\mathrel{\spose{\lower 3pt\hbox{$\mathchar"218$}}
     \raise 2.0pt\hbox{$\mathchar"13E$}}}
\def\simpropto{\mathrel{\spose{\lower 3pt\hbox{$\mathchar"218$}}
     \raise 2.0pt\hbox{$\propto$}}}

\def\ra{{\rm ra}}
\def\dec{{\rm dec}}

\def\h{{$^{\rm h}$}}
\def\d{{$^{\circ}$}}
\def\ll{{$<$}}

\def\ra{{$\alpha$}}
\def\dec{{$\delta$}}

\def\elle{{$\ell$}}

\documentclass[twocolumn,amsmath,nofootinbib]{revtex4} 
\usepackage{amsfonts,amsbsy,epsf} 
\begin{document}

\title{Clustering at 74 MHz}

\author{Ang{\'e}lica de Oliveira-Costa \& John Capodilupo}
\address{MIT Kavli Institute \& Dept.~of Physics, Massachusetts Institute of Technology, Cambridge, MA 02139}

\date{\today. To be submitted to MNRAS.}

\begin{abstract}
In order to construct accurate point sources simulations at the 
frequencies relevant to 21 cm experiments, the angular correlation 
of radio sources must be taken into account.
Using the 74 MHz VLSS survey, we measured the angular 2-point 
correlation function, $w(\theta)$. We obtain the first measurement 
of clustering at the low frequencies relevant to 21 cm tomography. 
We find that a single power law with shape $w(\theta) = {\rm A} 
\theta^{-\gamma}$ fits well the data. For a galactic cut of
$|b| >$ 10\d, with a data cut of $\delta >$ --10\d, and a flux
limit of $S$ = 770 mJy, we obtain a slope of 
	$\gamma$ = (--1.2$\pm$ 0.35).  
This value of $\gamma$ is consistent with that measured from other 
radio catalogues at the millimeter wavelengths. The amplitude of clustering 
has a length of 0.2\d -- 0.6\d, and it is independent of the 
flux-density threshold. 
\end{abstract}



\maketitle



\setcounter{footnote}{0}

\section{Introduction}

Progress in detector, space and computer technology has triggered 
an avalanche of high-quality cosmological data, removing cosmology 
from the realm of philosophy and transforming it into a quantitative 
empirical science. In the past few years, many authors have argued 
that the 21cm tomography, \ie, the three-dimensional mapping of highly 
redshifted 21cm emission, will be the ultimate cosmological probe -- 
see, \eg, 
	\cite{furlanetto02,zalda05a,zalda05b,furlanetto05,rennan04a,
	      babich05,rennan04b,asantha04,rennan05a,rennan05b,
	      BarkanaLoeb06}.
Although this signal has yet to be detected, there is a theoretical 
consensus that  the 21 cm signal must be out there and would be 
extremely useful if measured.
	\cite{Carilli04,furlanetto04,Carilli04b,miguel04,zaroubi04,
	      peterson05}

Although ambitious experimental efforts in 21cm tomography are now 
in progress across the globe (see \tabl{TomographyTable}), it is 
widely known/understood that the cosmological results of these 
experiments will only be as good as our ability to deal with (or 
to remove) foreground contamination 
	\cite{santos05,miguel05,wang,angelica,mao,liu1,liu2}. 
The goal of this work is to support these worldwide experimental 
efforts by tackling the foreground issue. 

Understanding the physical origin of Galactic metre wavelength emission 
is interesting for two reasons: to determine the fundamental properties 
of the Galactic components, and to refine the modeling of foreground 
emission for cosmological 21 cm experiments. 
At metre wavelengths, the bulk of foreground contamination is due 
to synchrotron emission. When coming from extragalactic objects, 
this radiation is usually referred to as point source contamination 
and affects mainly small angular scales. When coming from the Milky 
Way, this diffuse Galactic emission fluctuates mainly on large 
angular scales \cite{angelica}.

Normal galaxies, radio galaxies and active galactic nuclei form 
the majority of extragalactic continuum sources \cite{zotti09}. 
A number of surveys of radio sources have been performed at frequencies 
relevant to the 21 cm tomography -- see \tabl{mytable}; and analysis 
of these catalogs have helped to bring some understanding about their 
statistical properties: 
the distribution of radio sources is found to obey Poisson statistics 
with very weak observed angular clustering -- see \tabl{cooresults}.

\begin{figure}[pbt]\
\vskip-0.4cm
\centerline{\epsfxsize=7.5cm\epsffile{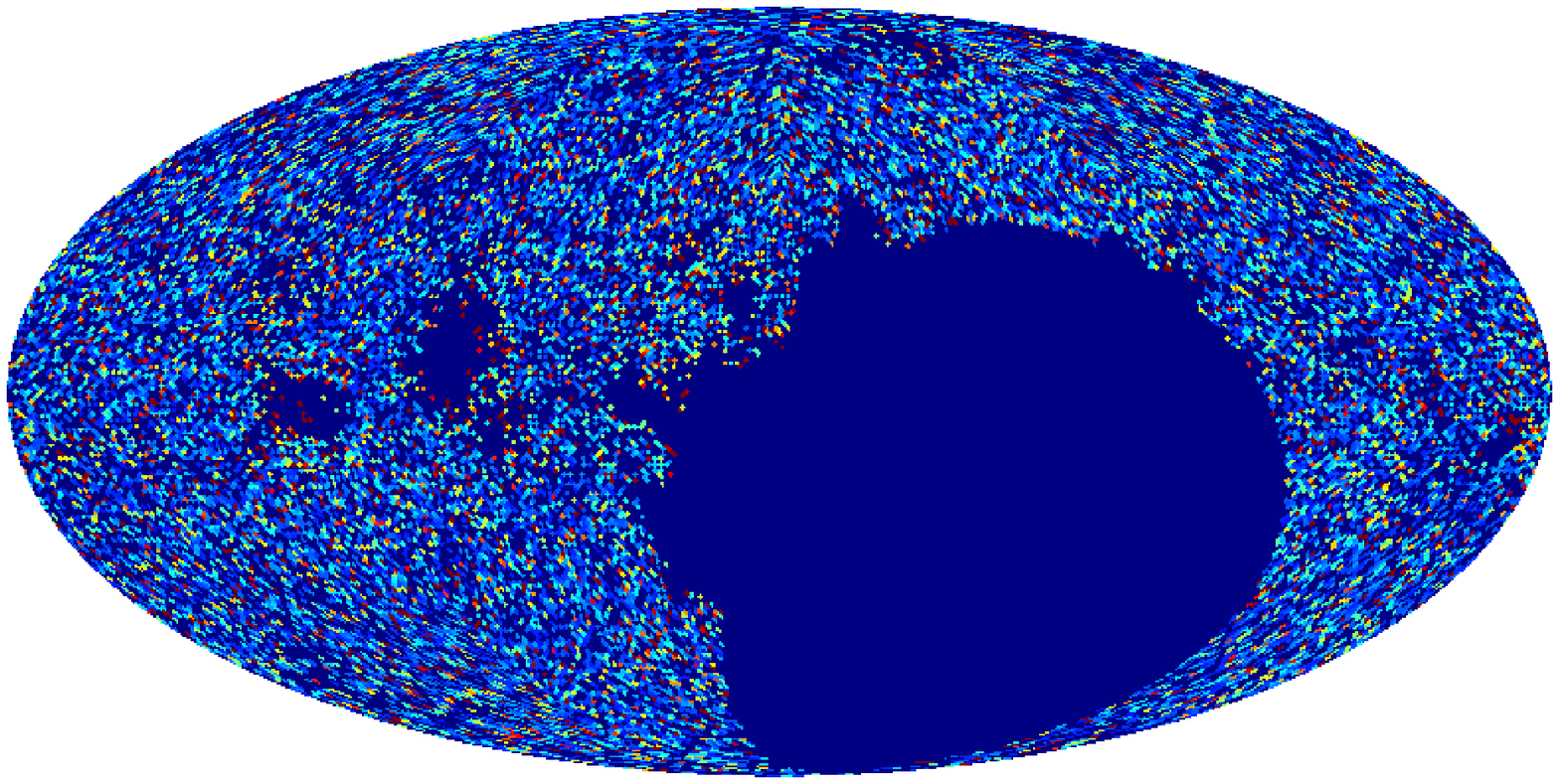}}
\centerline{\epsfxsize=7.5cm\epsffile{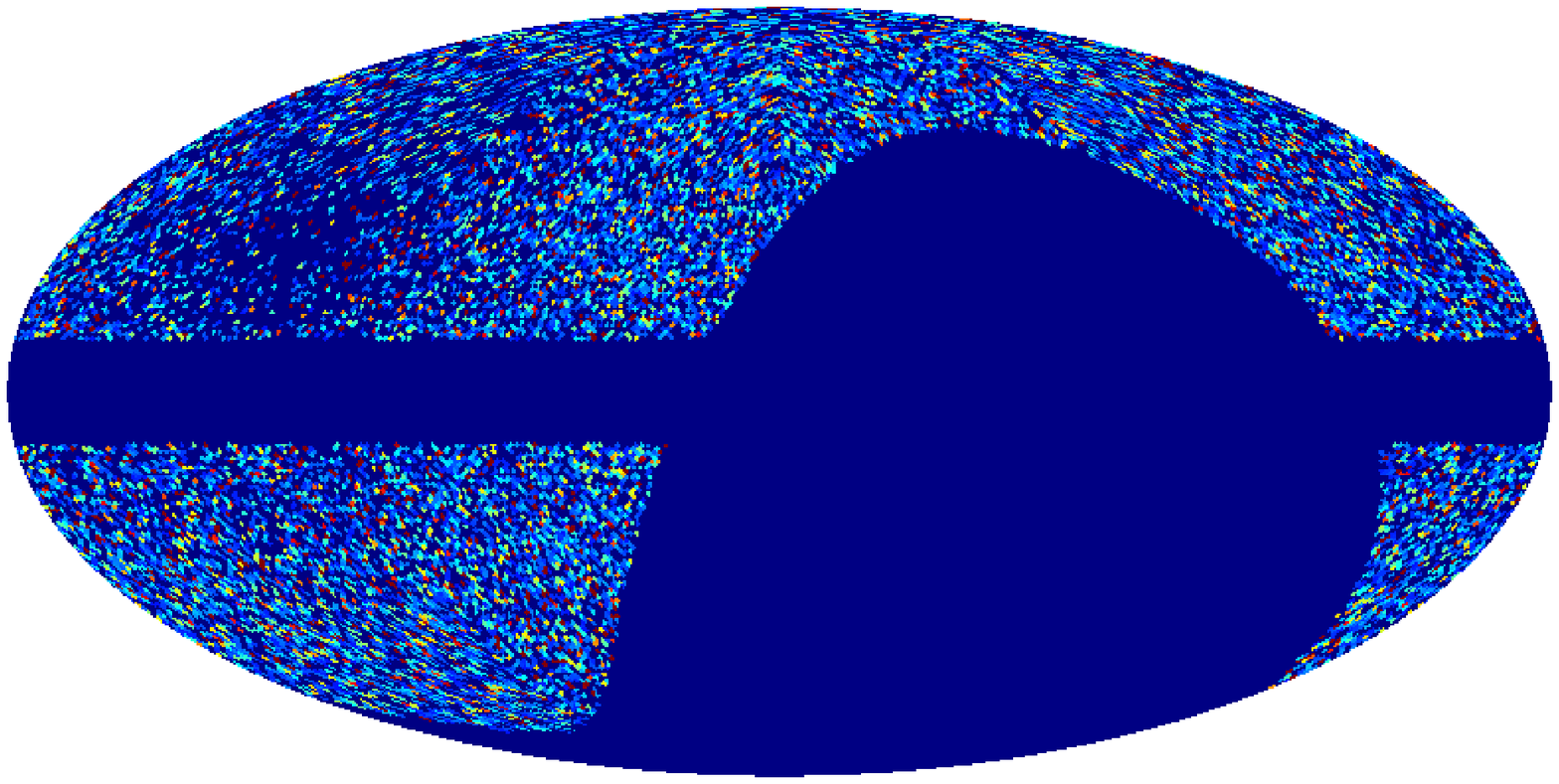}}
\vskip-0.3cm
\caption{\footnotesize%
         The 74 MHz VLSS catalogue (top), and one of our 74 MHz mock 
	 catalogues (bottom). Both catalogues are plotted in the interval 
	 of (0 $\le S \le 5$)Jy, and in Galactic coordinates with the Galactic 
	 center at the origin and longitude increasing to the left. The mock 
	 catalogue also shows the final area used in our analysis 
	 ($\delta$ $\ge$ --10\d and $|b| \ge$ 10\d).  
         } 
\label{4mass} 
\vskip-0.45cm
\end{figure}

Some aspects of both experimental design optimization and actual data 
analysis require full-blown simulations of the sky signal and 
knowledge about how it propagates through the instrument and the 
data analysis pipeline --- 
this has motivated the ambitious simulation efforts carried out by, 
\eg, WMAP and Planck. End-to-end simulations are at least as important 
for 21 cm experiments because of the many complicated issues related 
to instrumental performance, ionospheric turbulence corrections, 
{\etc} \cite{miguel03,bowman05,miguel05,judd,judd09}.
In order to construct accurate simulations at the metre wavelengths,
the angular correlation of radio sources must be taken into account 
\cite{gonzalez}. It is important to point out that the relative 
importance of the clustering contribution increases and may 
eventually become dominant if sources are identified and subtracted 
down to faint flux limits \cite{judd} -- which are exactly the limits 
involved in the point source removal of 21 cm experiments. 

\begin{table*}[!hr]
\caption{21 cm Tomography Experiments.} 
\medskip
\centerline{\label{TomographyTable}
\begin{tabular}{l|c|c|c|c|c|r}
\hline
\hline
\multicolumn{1}{l|}{Experiment}& 
\multicolumn{1}{|c|}{FWHM}& 
\multicolumn{1}{|c|}{$\nu$}&
\multicolumn{1}{|c|}{Receiver}& 
\multicolumn{1}{|c|}{Sensitivity}& 
\multicolumn{1}{|c|}{Effective Area}& 
\multicolumn{1}{|c}{Site-yr}\\
& &[MHz]&&&[m$^2$]&\\
  \hline 
  \hline
  GMRT        &3.8\d-0.4\d & 50--1420    &30 dishes		 &		       &5.10$^4$  &India	 - 2007\\
  PAST/21CMA  &3'          & 50--200     &10,000 antennas	 &15 mK/$\sqrt{day}$   &7.10$^4$  &Ulastai,CH    - 2007\\
  LOFAR       &25"-3.5"    & 10--240     &25,000 dipole antennas &		       &1.10$^5$  &Drenthe,NL	 - 2007\\
  MWA         &15'         & 80--300     &~8,192 dipole antennas &		       &1.10$^4$  &Murchison,AU  - 2007\\
  PAPER       &            & 110--200    &16 antennas		 &		       &1.10$^4$  &USA/AU	 - 2008\\
  SKA 	      &0.1"        & 100--25GHz  &			 &		       &	  &AU(?)	 - 2015(?)\\
  \hline
  \hline
\end{tabular} 
}
\smallskip
\noindent{\small \\
	         GMRT        = Giant Metrewave Radio Telescope, see http://www.gmrt.ncra.tifr.res.in/.\\
		 PaST/21CMA  = PrimevAl Structure Telescope, see http://web.phys.cmu.edu/~past/. \\ 
		 LOFAR       = LOw Frequency ARray, see http://www.lofar.org.\\
		 MWA         = Murchison Widefield Array, see http://www.haystack.mit.edu/ast/arrays/mwa/index.html. \\
		 PAPER       = Precision Array to Probe Epoch of Reionization, see http://astro.berkeley.edu/~dbacker/eor/. \\
		 SKA         = Square Kilometer Array, see http://www.skatelescope.org.
		 }\\
\end{table*}

\begin{table*}[!hr] 
\caption{Publicly available point source catalogues at the frequencies relevant to 21-cm tomography.} 
\medskip
\centerline{\label{mytable}
\begin{tabular}{l|c|cc|c|c|c|c|c|r}
\hline
\hline
\multicolumn{1}{ l|}{Ref}            &
\multicolumn{1}{|c|}{$\nu$}          &
\multicolumn{2}{ c }{Region}         &
\multicolumn{1}{|c|}{FWHM}           &
\multicolumn{1}{|c|}{S$_{comp}$}     &
\multicolumn{1}{|c|}{S$_{min}$}      &
\multicolumn{1}{|c|}{N$_{obj}$}      &
\multicolumn{1}{|c|}{Observatory}    &
\multicolumn{1}{|r }{Status}        \\
&[MHz] && &[arcmin] &[Jy] &[Jy] && \\
\hline
\hline 
 \protect\cite{8Cout}	       &  38	&00\h \ll \ra \ll 24\h  & +60\d \ll \dec \ll +90\d  &  4.5	&	& 1	 &5859       &CLFST, ENG      &A       \\  
 \protect\cite{Aslanyan}       &  60	&00\h \ll \ra \ll 24\h  & +55\d \ll \dec \ll +55\d  &450	&	&12	 &100	     &Pushchino, RUS  &B       \\  
 \protect\cite{4MASS}	       &  74	&00\h \ll \ra \ll 24\h  &--30\d \ll \dec \ll +90\d  &  1.33	&0.77	& 0.1	 &  68311    &VLA, USA        &A       \\  
 \protect\cite{Slee}	       &  80	&00\h \ll \ra \ll 24\h  &--49\d \ll \dec \ll +37\d  &  3.7	&	& 2	 & 999       &Culgoora, ENG   &A       \\  
 \protect\cite{Slee}	       &  80	&00\h \ll \ra \ll 24\h  &--49\d \ll \dec \ll +37\d  &  3.7	&	& 2	 & 1748      &Culgoora, ENG   &A       \\  
 \protect\cite{Branson}	       &  81	&00\h \ll \ra \ll 24\h  & +70\d \ll \dec \ll +90\d  & 10	&	& 1	 & 558       &Cambridge, ENG  &B       \\  
 \protect\cite{Dagkesamanskii} & 102 	&00\h \ll \ra \ll 24\h  & +27\d \ll \dec \ll +70\d  & 60	&	& 3	 & 920       &LPA, RUS        &A       \\  
 \protect\cite{MRT}	       & 150	&18\h \ll \ra \ll 24\h  &--70\d \ll \dec \ll--10\d  &  4.6	&2	& 0.96   & 2784      &MRT, India      &A       \\  
 \protect\cite{6Cout}	       & 151	&00\h \ll \ra \ll 24\h  & +30\d \ll \dec \ll +90\d  &  4.2	&	& 0.13   & 34418     &CLFST, ENG      &A       \\  
 \protect\cite{7Cout}	       & 151	&00\h \ll \ra \ll 24\h  & +21\d \ll \dec \ll +90\d  &  1.2	&	& 0.120  & 43689     &CLFST, ENG      &A       \\  
 \protect\cite{3Cout}	       & 159	&00\h \ll \ra \ll 24\h  &--22\d \ll \dec \ll +71\d  & 10.0	&	& 7	 & 471       &Cambridge, ENG  &A       \\  
 \protect\cite{Slee} 	       & 160	&00\h \ll \ra \ll 24\h  &--49\d \ll \dec \ll +37\d  &  1.85	&	& 1.2	 & 2041      &Culgoora, ENG   &A       \\  
 \protect\cite{3CRout}         & 178	&00\h \ll \ra \ll 24\h  &--90\d \ll \dec \ll--05\d  &  6.0	&	& 5	 & 11000     &Cambridge, ENG  &A       \\  
 \protect\cite{4Cout1,4Cout2}  & 178	&00\h \ll \ra \ll 24\h  &--07\d \ll \dec \ll +80\d  & 11.5	&	& 2	 & 4844      &4C Array, ENG   &A       \\  
 \protect\cite{MIYUN}	       & 232	&00\h \ll \ra \ll 24\h  & +30\d \ll \dec \ll +90\d  &  3.8	&	& 0.1	 & 34426     &MSRT, CHI       &A       \\  
 \protect\cite{WENSS}	       & 325	&00\h \ll \ra \ll 24\h  & +30\d \ll \dec \ll +90\d  &  0.9	&	& 0.1	 & 229420    &WSRT, NLD       &A       \\  
 \protect\cite{WISH}	       & 352	&00\h \ll \ra \ll 24\h  &--09\d \ll \dec \ll +26\d  &  0.9	&	& 0.010  & 84481     &WSRT, NLD       &A       \\  
 \protect\cite{TXS}	       & 365	&00\h \ll \ra \ll 24\h  & +36\d \ll \dec \ll +72\d  &  0.1	&	& 0.25   & 66841     &UTRAO, USA      &A       \\  
\hline								                        		
\hline
\end{tabular}
}
\smallskip
\noindent{\small \\
		 $S_{comp}$ = Limit of completeness. \\
		 $S_{min}$  = Smallest flux value. \\
		 $N_{obj}$  = Number of sources in the catalogue. \\
		 A = Publicly available in digital form.\\
		 B = Available as printed table (which we will OCR).  
		 }\\
\end{table*}

\clearpage

\begin{table}  
\caption{Published w$(\theta)^1$ values.} 
\medskip
\centerline{\label{cooresults}
\begin{tabular}{l|c|c|c|c|r}
\hline
\hline
\multicolumn{1}{ l|}{Ref}	     &
\multicolumn{1}{|c|}{$\nu$}	     &  
\multicolumn{1}{|c|}{A} 	     &
\multicolumn{1}{|c|}{$\gamma$}       &
\multicolumn{1}{|c|}{w$(\theta)$}    &
\multicolumn{1}{|c}{S$_{lim}$}      \\
&[GHz] &$\times 10^{-3}$ & &[\d]&[mJy]              \\	
\hline
\hline
 \protect\cite{seldner}  		&0.178	 &  	         &		  &1.50--3.0    &3000    \\
 \protect\cite{blake}          		&0.325	 &1.0 $\pm$ 0.4  &1.22 $\pm$0.33  &$>$ 0.2	&  35	 \\
 \protect\cite{webster77}               &0.408	 &  	         &		  &--	        & 250	 \\
 \protect\cite{webster77b}              &0.408	 &  	         &		  &--	        &  10	 \\
 \protect\cite{blake}          		&0.843	 &2.0 $\pm$ 0.4  &1.24 $\pm$0.16  &$>$ 0.2      &  10    \\
 \protect\cite{cress}       	        &1.400	 &2.6 $\pm$ 0.8  &1.2  $\pm$0.1   &0.07--4.0    &   3	 \\
 \protect\cite{magliocchetti}       	&1.400	 &1.1 $\pm$ 0.1  &1.5  $\pm$0.1   &$>$ 0.07     &   3	 \\
 \protect\cite{overzier}       		&1.400	 &1.0 $\pm$ 0.3  &0.9  $\pm$0.2   &$>$ 0.07     &   3	 \\
 \protect\cite{overzier,blake02a}	&1.400   &1.0 $\pm$ 0.2  &0.7  $\pm$0.1   &$>$ 0.1      &  10	 \\
 \protect\cite{blake}          		&1.400   &1.5 $\pm$ 0.2  &1.05 $\pm$0.10  &$>$ 0.3      &  10	 \\
 \protect\cite{webster77}               &2.7	 &  	         &		  &--	        & 350    \\
 \protect\cite{sicotte}                 &4.850   &  	         &		  &0.70--1.7    &  45    \\
 \protect\cite{kooiman}                 &4.850   &  		 &0.8 	          &0.30--1.9    &  35    \\
 \protect\cite{loan}                    &4.850   &10.0 $\pm$ 5.0 &0.8	          &0.01--1.0    &  50    \\
\hline
\hline
\end{tabular} 
}
\smallskip
\noindent{\small \\
		 $^1$w$(\theta)$ is fitted by a power-law of the form A$\theta^{-\gamma}$}\\
		 $S_{lim}$ = Smallest flux value.  
\end{table}

In this paper, we present measurements of the angular 2-point 
correlation function, $w(\theta)$, from the 74 MHz VLSS survey 
\cite{4MASS}. We obtain the first measurement of clustering 
at the low frequencies relevant to 21 cm tomography. In Section \ref{tools},
we described the statistical tools used in this analysis, as 
well as the 74 MHz VLSS survey. In Section \ref{results}, we describe
our results, and in Section \ref{discussion}, we present our conclusions.


\section{Data Analysis Tools}\label{tools}

\subsection{The Angular 2-point Correlation Function}

In recent years, the analysis of the correlation-function 
has become the standard way of quantifying the clustering 
of different populations of astronomical sources. 
Specifically, the angular two-point correlation function 
$w(\theta)$ gives the excess probability $\delta{\rm P}$, 
in comparison to a random Poisson distribution, of finding 
two sources in a solid angle $\delta\Omega_1$ and 
$\delta\Omega_2$ separated by the angle $\theta$. 
$\delta{\rm P}$ is defined as
\beq{prob}
	\delta{\rm P} = N^2 \delta\Omega_1 \delta\Omega_2
			 [ {\rm 1} + w(\theta)],
\eeq
where $N$ is the mean number density of objects in the 
catalogue under consideration \cite{Peebles80}. 

\begin{figure}[pbt]\
\centerline{\epsfxsize=9.1cm\epsffile{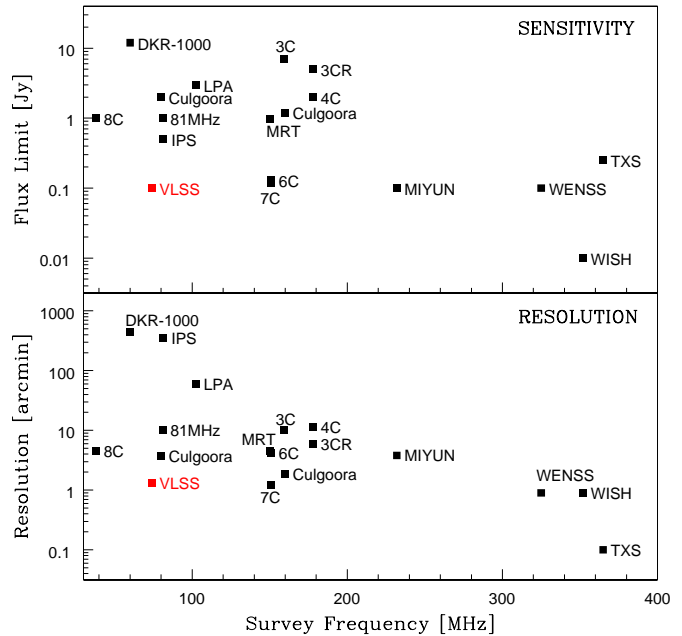}}
\vskip-0.4cm
\caption{A comparison between point source sensitivity and 
         resolution of the 74 MHz VLSS survey (in red) and 
	 other low frequency surveys (see 
	 \protect{\tabl{mytable}}).
         } 
\label{survey} 
\end{figure}

Many derivations for estimators of $w(\theta)$ can be found in 
the literature (see, \eg, \cite{Peebles80,LS93,H93}). One way 
to estimate this function is to compare the distribution of the
objects in the real catalogue to the distribution of points in 
a random Poisson distributed catalogue with the same boundaries, or 
\beq{w}
	w(\theta) = \frac{ DD(\theta) * RR(\theta)}{[DR(\theta)]^2} - 1 
\eeq
\cite{H93}, where $DD(\theta)$, $RR(\theta)$ and $DR(\theta)$ are the 
numbers of data-data, random-random and data-random pairs 
separated by the distance $\theta + \delta\theta$. 
It is important to remember that the estimation of $RR(\theta)$ 
and $DR(\theta)$ requires a catalogue of objects scattered 
uniformly over an area with the same angular boundaries of the 
data catalogue.

\subsection{Mock Catalogues}\label{mocks}

We used the ``Sphere Point Picking Algorithm" \cite{pick}
to generate random cartesian vectors equally distributed 
on the surface of a unit sphere (to avoid having vectors 
``bunched" around the poles, as it would happen if one 
chooses to plot the vectors in spherical coordinates instead). 
Accordingly, we calculate these vectors by doing
\beqa{ranvec}
	x = \sqrt{ 1 - u^2 } \cos\theta  \\
	y = \sqrt{ 1 - u^2 } \sin\theta  \\
	z = u, 			         
\eeqa  
where $u = \cos\phi$, with  $\theta \in [0,2\pi)$ and $u \in [-1,1]$ 
\cite{Weisstein}.
In order to obtain points such that any small area on the 
sphere is expected to contain the same number of points, 
we choose $u$ and $v$ to be random variates in the interval 
$[0,1]$. Therefore, we calculate $\theta$ and $\phi$ from
\beqa{thetaphi}
	\theta = 2 \pi u \\
	\phi = \cos^{-1} (2 v - 1).
\eeqa 

\begin{figure}[pbt]\
\vskip-0.3cm
\centerline{\epsfxsize=9.5cm\epsffile{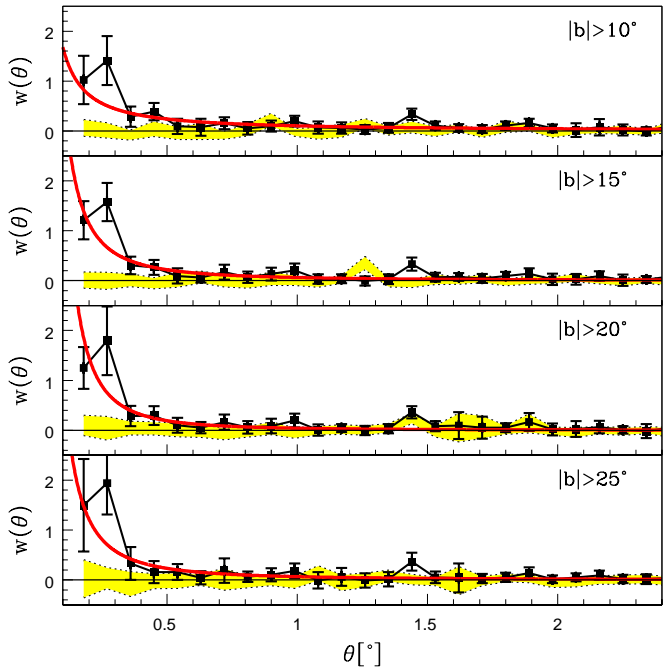}}
\vskip-0.4cm
\caption{Measured $w(\theta)$ for different galactic cuts.
	 All angular correlations are calculated at the 
	 flux limit of $S$ = 770 mJy. The red lines are single 
	 power law fits to the data, where 
	 	$w(\theta)$=A$\theta^{-\gamma}$;
	 and the yellow shaded regions are
	 $w(\theta)$ calculated using solely mocks. 
         } 
\label{corr} 
\end{figure}

Using the equations above we generate a position in the random 
catalogue. If this position is inside the boundaries of the data 
catalogue, then a temperature of the data catalogue is associated 
with that random vector. This procedure is repeated until the random 
catalogue has the same number of ``objects" as the data catalogue.
This method, also known as ``bootstrapping", involves resampling 
the data with replacement and, at random, to construct a new data 
set which has population distribution identical to that of the 
original dataset. \fig{4mass} shows a realization of one of our 
mock catalogues.

\subsection{VLSS: The VLA Low-Frequency Survey} 

The VLA Low-frequency Sky Survey (VLSS, formerly known as 4MASS)
is a 74 MHz (or 4 meter wavelength) continuum survey carried out 
by the National Radio Astronomy Observatory (NRAO) and the Naval 
Research Laboratory (NRL). The aim of the survey is to map an area 
of 3$\pi$ sr covering the entire sky north of $-30^{\circ}$ declination at 
resolution 80" (FWHM), with an average noise level of 0.1 Jy/beam. 
The principal data product is a set of 358 continuum images of 
($14^{\circ} \times 14^{\circ}$), and a catalogue with 68,311
discrete sources \cite{4MASS}.  
The VLSS catalogue was created by fitting elliptical Gaussians to 
all the sources that are detected at the 5 sigma level or higher 
\cite{lazio}, and it is complete at the 770 mJy level \cite{cohen4}.
The 74 MHz catalogue is shown in \fig{4mass}, top, and a comparison
of this survey with other low-frequency surveys can be seen
in \tabl{mytable} and \fig{survey}.  


\section{Results}\label{results}

In \fig{corr}, we present our measurement of $w(\theta)$ 
for the flux limit of $S$ = 770 mJy (black squares), which 
is the completeness limit of the VLSS catalogue. Distances 
between data and/or random sources are measured in bins of 
0.09$^\circ$, which is safely above the VLSS resolution 
limit of 0.02$^\circ$. We also investigated if $w(\theta)$ 
changes with bin size, and we found no indication that 
any change in bin size affects our results. 

\begin{figure}[pbt]\
\vskip-4.0cm
\centerline{\epsfxsize=8.8cm\epsffile{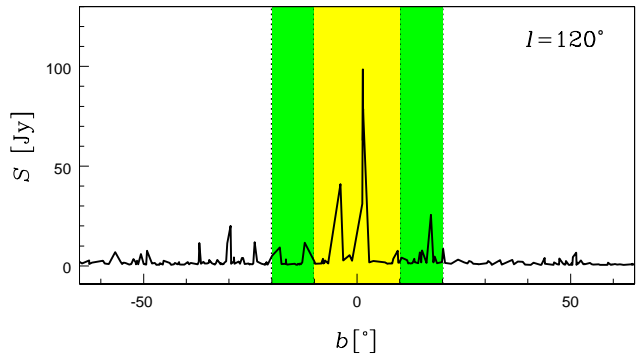}}
\vskip-0.4cm
\caption{The signal as a function of galactic latitude for a
         constant galactic longitude of \elle=120\d. Green
	 and yellow shades enclose the regions $|b| \le$ 20\d 
	 and $|b| \le$ 10\d, respectively.
	 Sources within these shaded regions are masked from 
	 our analysis as they may be galactic in origin.
         } 
\label{galcut} 
\end{figure}

As shown in \fig{galcut}, there are sources in the VLSS 
catalogue that may be galactic in origin. In this figure, 
we plot the source fluxes at galactic longitude $\ell$=120\d  
as a function of galactic latitude $b$. The green and yellow 
shades enclose the regions $|b| \le$ 20\d and $|b| \le$ 10\d, 
respectively. To reduce contamination from galactic sources, 
we discarded regions inside chosen Galactic cuts; we also 
discarded regions below $\delta < $ --10\d, due to the patch 
sky coverage of VLSS -- see \fig{4mass}, top. 

From top-to-bottom, \fig{corr} shows the measured 
$w(\theta)$ for different galactic cuts. We detected 
no correlation for cuts smaller than 10$^\circ$ and, above 
this limit, there are no large variations in $w(\theta)$. 
Since we want to maximize the number of sources used in our 
statistics, from here on, all final calculations are for
a 10$^\circ$ galactic cut (\ie, for 39,118 sources). 
	A galactic cut of 10$^\circ$ (or bigger) also excludes
	the ``blank" regions in the VLSS survey --
	see \fig{4mass}. They are regions around, 
	\eg, CasA and Cyg A. 

We construct 100 mock catalogues using the procedure described 
in \ref{mocks}, with flux values above the sensitivity limit 
of the data catalogue and a chosen galactic cut of 10$^\circ$ 
applied. 
By cross-correlating the data with the 100 mocks, we produce 
a set of normally distributed estimates of the correlation 
function. The mean and the standard deviation of this 
distribution are used as a value for the estimate and its 
uncertainty in the measurement of $w(\theta)$ at each 
$\theta$\footnote{Data points in a plot of $w(\theta)$ are 
        not independent, \ie, single sources can contribute 
	pairs in more than one bin. Therefore, standard Poisson 
	error bars will underestimate the true error in each 
	bin. 
       }.
The estimate (mean) and its uncertainty (the standard 
deviation) are shown in \fig{corr} as the black 
squares and their error bars. 
Similarly, we correlated the 100 mocks with themselves.  
This result correspond to the yellow shaded region shown 
in \fig{corr} and, as expected in a Poissonian distribution, 
$w(\theta)$ is consistent with zero.

\begin{table}  
\caption{w$(\theta)^1$ results.} 
\medskip
\centerline{\label{mycooresults}
\begin{tabular}{l|c|c|c|c|r}
\hline
\hline
\multicolumn{1}{ l|}{$|b|$}	     &
\multicolumn{1}{|c|}{A} 	     &
\multicolumn{1}{|c|}{$\gamma$}       &
\multicolumn{1}{|c|}{w$(\theta)$}    &
\multicolumn{1}{|c|}{S$_{lim}$}      &
\multicolumn{1}{|c }{$\chi^2$}      \\
& & &[\d]&[mJy]&                    \\	 
\hline
\hline
  10\d        &0.103$\pm$0.026    &-1.21$\pm$0.35     &0.2--0.6   &770	&0.62	 \\
  15\d        &0.062$\pm$0.011    &-1.81$\pm$0.47     &0.2--0.6   &770	&0.73	 \\
  20\d        &0.041$\pm$0.007    &-2.22$\pm$0.78     &0.2--0.6   &770	&0.57	 \\
  25\d        &0.066$\pm$0.011    &-1.81$\pm$0.28     &0.2--0.5   &770	&0.58	 \\
\hline  
  10\d        &0.113$\pm$0.029    &-1.09$\pm$0.20     &0.2--0.6   &850	&0.86	 \\
  10\d        &0.104$\pm$0.028    &-1.26$\pm$0.38     &0.2--0.6   &900	&0.63	 \\
\hline
\hline
\end{tabular} 
}
\smallskip
\noindent{\small \\
		 $S_{lim}$ = Smallest flux value. \\ 
		 $^1$w$(\theta)$ is fitted by a power-law of the form A$\theta^{-\gamma}$} 
\bigskip
\end{table}
%
%
%
%

\begin{figure}[pbt]\
\vskip-4.0cm
\centerline{\epsfxsize=9.3cm\epsffile{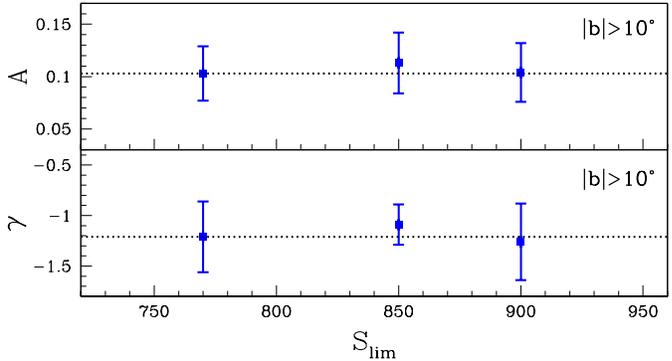}}
\vskip-0.4cm
\caption{Measured amplitudes of A and $\gamma$ for various 
         flux-density limits at 770 mJy, 850 mJy and 900 mJy.
	 Note that the amplitude of clustering does not depend 
         on flux density.
         } 
\label{AGamma} 
\end{figure}

We find that a single power law with shape 
    $w(\theta) = {\rm A} \theta^{-\gamma}$ \cite{Peebles80}, 
where A is a measure of the amplitude of the average 
enhancement of the number of radio sources at a particular 
point in the sky, fits the data well. We present our 
measurements in \tabl{mycooresults}.
We also calculate $w(\theta)$ for various flux-density limits 
at 770 mJy, 850 mJy and 900 mJy. As shown in \tabl{mycooresults}
and \fig{AGamma}, the amplitude of clustering does not depend 
on flux density. This same result was observed  
in previous angular correlation analysis (\eg, \cite{blake}).


Some other interesting results can be taken from \fig{corr}:
(1)
at large angular separations, $\theta >> $ 2\d, $w(\theta)$ is 
consistent with zero --  this is a strong evidence for a 
high degree of uniformity in the survey. 
(2)
at the small angular separations, $\theta < $ 0.2\d, there 
is a fall-off (or a break) in the value of $w(\theta)$ -- 
this effect is due to the failure of the survey to resolve 
weak double sources with separations slightly greater than 
the beamwidth. \cite{blake02b} presents a detailed explanation 
of this effect and show, in details, how this calculation 
is done. 
(3)
Finally, at the angular separation of $\theta \approx $1.45\d, 
there is an unexplained increase in the value of $w(\theta)$. 
It is well studied and reported in the literature that instrumental 
effects in radio surveys manifest themselves on particular 
characteristic scales, and are usually rendered transparent 
by the $w(\theta)$ analysis (see, \eg. \cite{blake02b}). 
If the anomaly described above is caused by such effects, this is
something that should be carefully studied, but it is outside the 
scope of this paper. 


\section{Discussion}\label{discussion}

In order to construct accurate simulations at the metre wavelengths,
the angular correlation of radio sources must be taken into account.
The relative importance of the clustering contribution increases and 
may eventually become dominant if sources are identified and subtracted 
down to faint flux limits -- which are exactly the limits involved 
in the point source removal of 21 cm experiments. 

Using the 74 MHz VLSS survey, we measured the angular 2-point 
correlation function, $w(\theta)$. We obtain the first measurement 
of clustering at the low frequencies relevant to 21 cm tomography. 
We find that a single power law with shape $w(\theta) = {\rm A} 
\theta^{-\gamma}$ fits the data well. For a galactic cut of
$|b| >$ 10\d, with a data cut of $\delta >$ --10\d, and a flux
limit of $S$ = 770 mJy, we obtain a slope of 
	$\gamma$ = (--1.2$\pm$ 0.35) 
with $\chi^2$=0.62. This value of $\gamma$ is consistent with that 
measured from other radio catalogues -- see \tabl{cooresults}. 
The amplitude of clustering has a length of 0.2\d -- 0.6\d, and it
is independent of the flux-density threshold. 

\bigskip
\bigskip
\bigskip

\noindent
{\bf ACKNOWLEDGMENTS:}
\\

\noindent The authors wish to thank Joseph Lazio and Mike Matejek 
for helpful comments. Support for this work was provided by NSF 
through grants AST-0607597 and AST-0908950. JC acknowledges the 
Center for Excellence in Education for holding the Research 
Science Intitute (RSI) at MIT to support this work.


\clearpage


\end{document}